\documentclass[amsmath,amssymb,aps,prd,nofootinbib]{revtex4-2}
\usepackage{amsmath}
\usepackage[normalem]{ulem}
\usepackage{color}
\usepackage{graphicx}
\usepackage{epsfig}
\usepackage{subfig}
\usepackage{bm}
\usepackage{algorithm}
\usepackage{hyperref}
\usepackage{natbib}
\usepackage{pgfplots,mathtools}
\usepackage{braket}
\usepackage{slashed}
\usepackage[compat=1.0.0]{tikz-feynman}
\usepackage{physics}
\usepackage{mathrsfs}
\usepackage[utf8]{inputenc}
\usepackage{academicons}
\newcommand{\orcid}[1]{\href{https://orcid.org/#1}{\includegraphics[width=8pt]{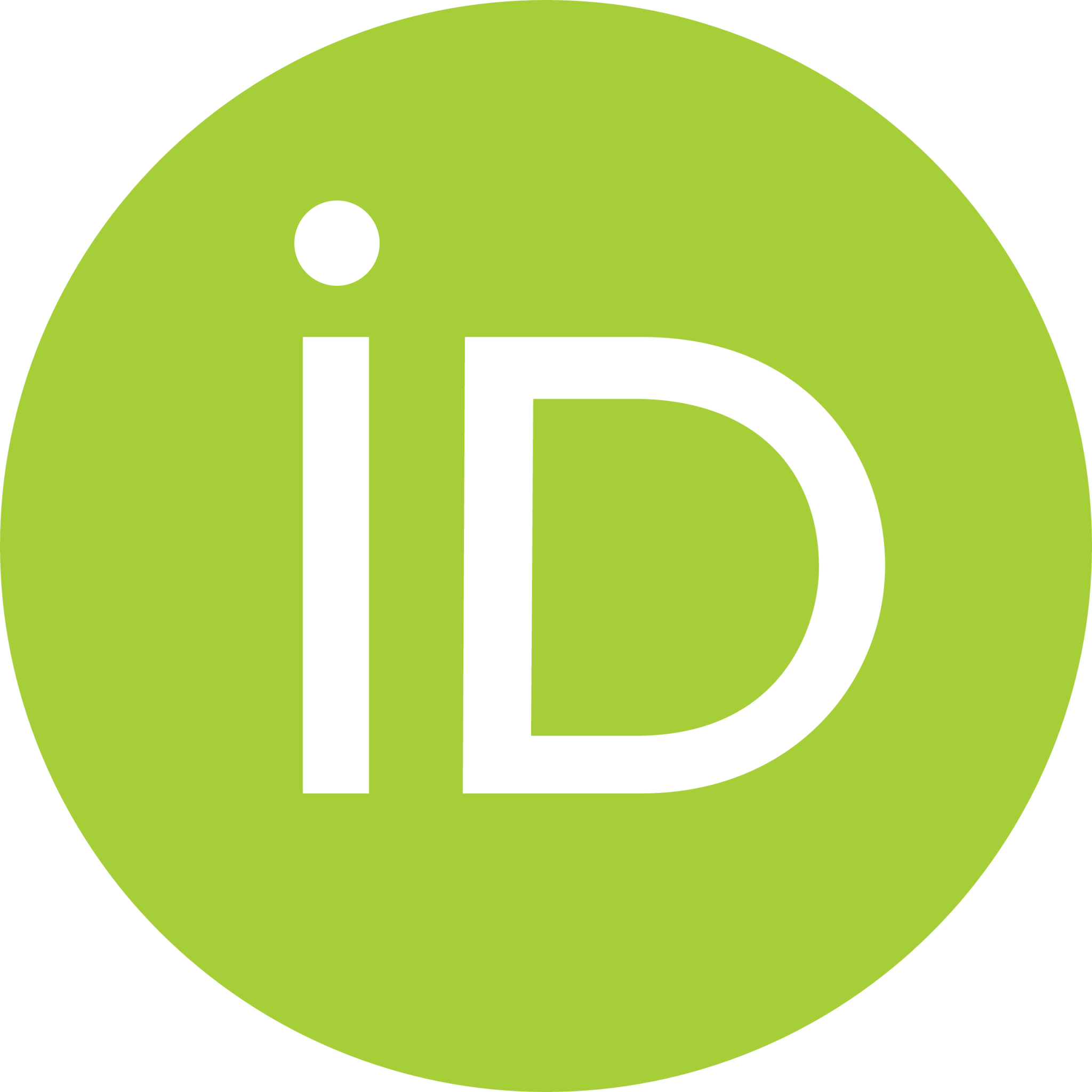}}}
\usepackage{xfrac}
\usepackage{xcolor}

\begin{document}
\title{Possibility of quantum Hall effect in dense quark matter environments: A chiral model approach}
\author{Dani Rose J Marattukalam\orcid{0009-0006-8204-8148}, Ashutosh Dwibedi\orcid{0009-0004-1568-2806}, Sourodeep De\orcid{0000-0003-1411-9486} and Sabyasachi Ghosh\orcid{0000-0003-1212-824X}}
\affiliation{Department of Physics, Indian Institute of Technology Bhilai, Kutelabhata, Durg, 491002, Chhattisgarh, India }
%

%
\begin{abstract}
	A high baryon density and strong magnetic fields are expected in peripheral collisions in heavy-ion collision experiments, such as the upcoming CBM experiment at FAIR in Germany and NICA in Russia. Such densities are also likely in the core of massive neutron stars, possibly with mixed quark-hadron phases. We employed the chiral effective model to obtain the constituent quark mass in this non-perturbative QCD regime. A quantized version of conductivity and resistivity is found reliable in the quantum domain of low density and high magnetic fields. Landau quantization gives rise to phenomena similar to SdH oscillations and quantum Hall effect in this regime. We have used a density-dependent magnetic field to observe SdH-type oscillations and the possibility of the quantum Hall effect in the interior of neutron stars where the magnetic field varies as a function of the baryon density. Our results indicate the possibility of observing the quantum Hall effect in a neutron star environment.
\end{abstract}
\maketitle
\section{Introduction}
The study of stellar remnants formed in the aftermath of supernova explosions like the white dwarfs, neutron stars and other astrophysical objects --- which are subject to conditions of high densities and magnetic fields~\cite{Harding:2006qn} --- has been an active area of research and has attracted the attention of astrophysicists and theoretical physicists alike. Various studies have shown the strength of the magnetic field on the surface of the neutron star to be of the order of $10^{12}$G to $10^{15}$G ~\cite{Seiradakis:2004ew}.
Due to intense pressure in the interior of neutron/hybrid/quark stars, the supposed hadron matter can undergo deconfinement due to asymptotic freedom 
~\cite{Khadkikar:1995gh}. Attempts have been made to obtain the equation of state (EOS) of neutron stars and other compact objects using effective field theories ~\cite{Broderick:2000pe,Dexheimer:2008ax,Hanauske:1999ga,Mishra:2003tr,Zschiesche:2003qq,Mishra:2006wy} where the behaviour of the nuclear matter is modelled by the application of scalar attractive and vectorial repulsive mesons. It is their interplay which dictates the EOS of these astrophysical objects. 

In the context of the QCD phase diagram, systems like the atomic nuclei and neutron stars correspond to conditions of low temperatures and high density and are often referred to as cold nuclear matter. On the other end of the spectrum are the hot and dense phases, such as the quark-gluon plasma that existed in the early universe just after the Big Bang. Ultra-relativistic heavy-ion collision experiments at the Large Hadron Collider (LHC) at CERN and  Relativistic Heavy Ion Collider (RHIC) in BNL probe matter at high temperatures and nearly vanishing net baryon densities~\cite{Braun-Munzinger:2015hba,Elfner:2022iae}, while the upcoming facilities of Compressed Baryonic Matter (CBM) experiment in FAIR and Nuclotron-based Ion Collider fAcility (NICA) at JINR will probe the highest baryonic densities ever obtained in laboratory environments~\cite{CBM:2016kpk,Senger:2020pzs,Senger:2020wvj,Sissakian:2009zza} at which mixed quark-hadron phases are expected. 

Observables like the charge-dependent directed flow in heavy ion collisions ~\cite{STAR:2023jdd} provide hints of high magnetic fields and can constrain electrical conductivity calculations. 
Throughout the last decade, numerous research ~\cite{Chatterjee:2019nld,Ghosh:2019ubc,Ghosh:2022xtv,Ghosh:2020wqx,Dash:2020vxk,Dey:2019vkn,Huang:2011dc,Ghosh:2018cxb,Dey:2019axu,Dey:2019vkn,Satapathy:2019jdw,Kalikotay:2020snc,Singh:2023ues,Goswami:2023eol,Rath:2020idp,Rath:2021ryd,Ghosh:2018cxb,Dey:2019axu,Dey:2019vkn,Satapathy:2019jdw,Kalikotay:2020snc,Singh:2023ues,Goswami:2023eol,Rath:2020idp,Rath:2021ryd,Shaikh:2024gsm,Satapathy:2021cjp,Satapathy:2021wex,Ghosh:2024owm,Ghosh:2024fkg,Das:2019wjg,Das:2019pqd,Dash:2020vxk,Panda:2020zhr,Panda:2021pvq,Kushwah:2024zgd,Sen:2021tdu,sedrakyan1987magnetohydrodynamics,Potekhin:1999ur,Chamel:2008ca,Harutyunyan:2016rxm,Harutyunyan:2023ooz,Harutyunyan:2024hsd,Buividovich:2010tn,Buividovich:2010qe,Dey:2020awu,Dey:2021fbo,Bandyopadhyay:2023lvk,Satapathy:2022xdw} was done to understand the properties of nuclear matter at finite magnetic fields motivated by the possibility of high magnetic field production in RHIC ($eB \sim 1~m_\pi^2$) and LHC ($eB \sim 10~m_\pi^2$) experiments \cite{Tuchin:2013ie}. Despite being an older subject, the study of finite and high magnetic fields in neutron star environments ~\cite{sedrakyan1987magnetohydrodynamics,Potekhin:1999ur,Chamel:2008ca} still attracts considerable scientific interest ~\cite{Harutyunyan:2016rxm,Harutyunyan:2023ooz,Harutyunyan:2024hsd}. The topic of dense QCD matter at finite magnetic fields, characterized by low temperature and high density, is gaining momentum in the context of future experimental facilities at CBM and NICA. As far as RHIC or LHC matter is concerned, we find two well-practised methods for the calculation of transport coefficients --- the relaxation time approximation in kinetic theory ~\cite{Dey:2019vkn,Ghosh:2018cxb,Dey:2019axu,Dey:2019vkn,Satapathy:2019jdw,Kalikotay:2020snc,Singh:2023ues,Goswami:2023eol,Rath:2020idp,Rath:2021ryd,Ghosh:2018cxb,Dey:2019axu,Dey:2019vkn,Satapathy:2019jdw,Kalikotay:2020snc,Singh:2023ues,Goswami:2023eol,Rath:2020idp,Rath:2021ryd,Shaikh:2024gsm} and the diagrammatic Kubo-type formalisms ~\cite{Ghosh:2019ubc,Satapathy:2021cjp,Satapathy:2021wex,Ghosh:2024fkg,Ghosh:2024owm}. Calculations including the HRG model ~\cite{Das:2019wjg,Das:2019pqd,Dash:2020vxk,Rocha:2024rce}, NJL model ~\cite{Ghosh:2019ubc,Islam:2019tlo}, quasiparticle approach~\cite{Das:2019ppb} and magnetohydrodynamics~\cite{Panda:2020zhr,Panda:2021pvq,Kushwah:2024zgd} are  documented.  Finite density calculations considering quark and hadronic phases in neutron stars are carried out in~\cite{Sen:2021tdu}. Magnetohydrodynamics description of plasma in the neutron stars~\cite{sedrakyan1987magnetohydrodynamics} and transport coefficients, including electrical conductivity ~\cite{Potekhin:1999ur,Chamel:2008ca,Harutyunyan:2016rxm,Harutyunyan:2023ooz} and thermal conductivity~\cite{Harutyunyan:2024hsd}, were always topics of interest. Refs. ~\cite{Buividovich:2010tn,Buividovich:2010qe} use lattice simulations in the quenched SU(2)
theory to study the effect of external magnetic fields on electrical conductivity. 

However, in the strong magnetic field limit, the transport coefficients are quantized  ~\cite{Dey:2020awu,Dey:2021fbo,Bandyopadhyay:2023lvk,Satapathy:2022xdw}. Two interesting macroscopic manifestations of Landau quantization are the Shubnikov-de Haas (SdH) oscillations and the Quantum Hall Effect (QHE). This quantization is expected to manifest in observables in the neutron star environment.  The possibility of observation of the Quantum Hall Effect in neutron stars where the Hall component of conductivity and resistivity is quantized into distinct plateaus was discussed in ~\cite{Dey:2021fbo}, considering massless quark matter. Calculations show periodic oscillations in the electrical conductivity of quarks due to Landau quantization similar to the Shubnikov-de Haas (SdH) oscillations observed in electronic systems at high magnetic fields and low temperatures. Similar estimates of electrical conductivity due to degenerate electrons in the outer envelopes of magnetized neutron stars are reported in~\cite{Potekhin:1999ur}. The present work intends to explore the possibility of these SdH/QHE oscillations in low temperature, high magnetic field conditions present in the interior of the neutron star and the upcoming CBM/NICA experiments. A net baryon density up to 2-6 times the nuclear saturation density is expected in the reaction zone of the nuclear collisions at CBM energies~\cite{Senger:2020wvj}. Such densities are also expected in the core of massive neutron stars~\cite{Lattimer:2021emm}. Lattice QCD calculations are restricted to the low density region due to the numerical sign problem and one needs to either use effective QCD models in the whole density range or the perturbative QCD (pQCD) at asymptotically large densities.  As a matter of fact, the phase structure of the nuclear matter with moderate to high baryon density and zero temperature has been under intense investigation for the last two decades (See~\cite{RevModPhys.80.1455,RevModPhys.86.509} and references therein). For instance, in cold quark matter, the Cooper pairing can lead to gaps in the energy spectrum of the quasiparticle excitations near the Fermi surface~\cite{Bailin:1983bm,IWASAKI1995163,Alford:1997zt,Rapp:1997zu,Berges:1998rc}. At negligible temperature and asymptotically high density (say, $\rho\gtrsim 40\rho_{0}$), the most favored phase of quark matter is the Color Flavor Locked (CFL) phase, where the fermionic quasiparticle excitations are gapped, and the transport is carried out by Goldstone modes~\cite{Alford:1998mk,Rajagopal:1998ec}. In the moderately high density $\rho \approx 2-4\rho_0$ or $\rho \leq 4\rho_0$, various phases of the dense quark matter, viz, two flavor superconductors (2SC), crystalline superconductors, etc ~\cite{RevModPhys.80.1455,RevModPhys.86.509,Buballa:2003qv,Mishra:2022pee}, has been postulated because of the mismatch of the constituent masses of different quark flavors. Gapless fermionic excitations contributing to the transport phenomena can appear in moderate baryon density scenarios for phases like 2SC. A kinetic theory approach to transport coefficients in the 2SC phase of quark matter can be found in Ref. \cite{Sarkar:2016gib}. In this respect, the kinetic theory framework for quark transport coefficients presented in this work is directly applicable only within a limited range of temperature and density where quarks remain in the normal (non-superconducting) phase with gapless dispersion relations. We also do not consider effects such as the inhomogeneous chiral symmetry breaking \cite{Basar:2010zd,Basar:2012gm,Basar:2009fg,Basar:2008im,Basar:2008ki}. In this work, to represent neutron star conditions, we consider a temperature $T=1$ MeV, and for dense QCD matter relevant to future CBM or NICA experiments, we assume $T=100$ MeV. Electrical conductivity and resistivity tensors are sketched as functions of density from $\rho=0$ to $\rho=4\rho_0$ at these fixed temperatures. However, only the density range corresponding to gapless quarks in the normal phase is physically relevant. Within this domain, we highlight the possible quantization patterns of conductivity/resistivity in neutron star environments, which are expected to diminish as one approaches the CBM/NICA regimes. To map this increasing baryon density and mixed quark-hadron phases we employed the quasiparticle formalism with constituent quarks as the quasiparticles of the system. In the hadronic phase, we have considered quarks confined within the protons to be degrees of freedom by using the chiral effective model to account for the constituent quark mass. This helps us use a single formalism over the whole range of densities rather than considering hadronic and deconfined quark phases separately. 

One can use different effective QCD models for moderately high (baryon) density or chemical potential. One such effective QCD model is the chiral model~\cite{De:2022zka,De:2022gse, Papazoglou:1997uw, Papazoglou:1998vr} adopted in the present work. According to this model, the quark condensate and constituent quark mass decreases as density or chemical potential increases. At huge (baryon) density or chemical potential, the condensate is expected to melt down to restore chiral symmetry, and
the constituent quark mass converts to the current quark mass where perturbative QCD (pQCD) calculations are valid.
Thus, the chiral model provides a non-pQCD estimation along the intermediate density domain where
both LQCD and pQCD techniques fail. This work aims to provide a realistic, non-pQCD estimation of the electrical conductivity and resistivity components in the presence of a magnetic field, building on preliminary estimates made in~\cite{Dey:2021fbo} using a simple, density-independent quark mass.

The rest of the paper is organized as follows: Section \ref{IIA} describes the chiral effective model and the calculation of constituent quark mass from the quark condensates. In Section \ref{IIB}, we discuss the classical and quantum expressions of electrical conductivity and resistivity in the relaxation time approximation of kinetic theory. In Section \ref{III}, we discuss in detail our results on the electrical conductivity and resistivity of a degenerate quark system described in terms of both the constituent and current quark masses. We also present a more realistic estimate of the same in the interior of a neutron star where the magnetic field varies with the baryon density. Estimates of electrical conductivity for dense quark matter at a higher temperature relevant to future experimental facilities at CBM/NICA are also presented. We summarize our study in Section \ref{IV}. 

\section{Formalism}
 \subsection{Constituent quark mass from chiral effective model}\label{IIA}
In this section, we describe and use the chiral effective model to obtain the expressions for constituent quark masses. The chiral model is based on the principles of chiral symmetry breaking and the broken scale invariance of QCD. In this model, we start with a Lagrangian density, having kinetic part, $\mathscr{L}_{kin}$ and other parts that include the baryon - scalar meson interaction term $\mathscr{L}_{BX}$, 
the spontaneous chiral symmetry breaking term given by meson - meson interaction $\mathscr{L}_{SSB}$, the QCD scale invariance breaking term $\mathscr{L}_{scale-break}$, and the explicit chiral symmetry breaking term $\mathscr{L}_{ESB}$ \cite{Mishra:2003tr, Zschiesche:2003qq, Mishra:2006wy,De:2022zka,De:2022gse, Papazoglou:1997uw, Papazoglou:1998vr, Parui:2022msu}. The effective Lagrangian $\mathscr{L}$ thus contains the following terms: 
\begin{equation}\label{L_density}
	\mathscr{L} = \mathscr{L}_{kin}+ \mathscr{L}_{BX} +\mathscr{L}_{SSB} + \mathscr{L}_{scale-break} + \mathscr{L}_{ESB}. 
\end{equation}
where,
\begin{eqnarray}
	\mathscr{L}_{BX} &=& -\sum_{i=n,p} \bar{\psi_i} m_i^*\psi_i,\\
	\mathscr{L}_{SSB} &=& -\frac{1}{2}k_0\chi^2(\sigma^2 + \zeta^2 + \delta^2) + k_1(\sigma^2 + \zeta^2 + \delta^2)^2 +  k_2\Bigg(\frac{\sigma^4}{2} + \frac{\delta^4}{2} + 3\sigma^2\delta^2 + \zeta^4\Bigg)\nonumber \\
	&+& k_3\chi(\sigma^2 - \delta^2)\zeta - k_4\chi^4,\\
	\mathscr{L}_{scale-break}&=& - \frac{1}{4}\chi^4 \ln{\frac{\chi^4}{\chi_0^4}} + \frac{d}{3}\chi^4 \ln{\Bigg( 
		\frac{(\sigma^2 - \delta^2)\zeta}{\sigma_0^2\zeta_0} \Big(\frac{\chi}{\chi_0}\Big)^3   \Bigg)},
	\label{sb}\\
	\mathscr{L}_{ESB} &=& -\Big(\frac{\chi}{\chi_0}\Big)^2 Tr\Bigg[diag \Bigg(\frac{1}{2}m_\pi^2f_\pi(\sigma + \delta),~ \frac{1}{2}m_\pi^2f_\pi(\sigma - \delta),~ \Big(\sqrt{2}m_K^2f_K - \frac{1}{\sqrt{2}}m_\pi^2f_\pi \Big) \zeta \Bigg)\Bigg].\label{L_ESB}
 \end{eqnarray}
Here, 
\begin{equation}
	m^*_i = -g_{\sigma i} \sigma - g_{\zeta i}\zeta - g_{\delta i}\delta
\end{equation}
is the effective mass of the baryon field $\psi_{i}$. Since we are considering nuclear matter, $i=p,n$ stand for proton and neutron, respectively. $\sigma, \delta$ and $\zeta$ are the scalar fields, and $\chi$ is the dilaton field. The values of the baryon-meson couplings are given by $g_{\sigma p} = g_{\sigma n} = 10.567$,  $g_{\zeta p} = g_{\zeta n} = -0.467$,  
and the values of decay constants and masses of mesons are $f_\pi = 93.3$ MeV, $f_K = 122.143$ MeV, $m_\pi = 139$ MeV and $m_K = 498$ MeV.

Comparing Eq.~\eqref{L_ESB} in the frozen glue-ball limit ($\chi=\chi_0$) with the explicit chiral symmetry breaking term in the QCD Lagrangian,
\begin{equation}
	\mathscr{L}_{ESB}^{QCD} = -Tr\bigg[diag(m_u \Bar{u}u,~m_d \Bar{d}d,~ m_s \Bar{s}s)\bigg],
\end{equation}
we have,
\begin{align}
	m_u\langle \Bar{u}u \rangle = \frac{1}{2}m_\pi^2f_\pi(\sigma+\delta)\label{uu_bar},\\
	m_d\langle \Bar{d}d \rangle = \frac{1}{2}m_\pi^2f_\pi(\sigma-\delta)\label{dd_bar}, \\
	m_s\langle \Bar{s}s \rangle = \Bigg(\sqrt{2}m_K^2f_K - \frac{1}{\sqrt{2}}m_\pi^2f_\pi \Bigg)\zeta\label{ss_bar}.
\end{align}
The  equations of motion for the fields $\sigma, \zeta, \delta$ and $\chi$, obtained from  the Lagrangian density in Eq.~\eqref{L_density}, are ~\cite{Mishra:2003tr,Mishra:2006wy,De:2022zka,De:2022gse,Parui:2022msu}
\begin{multline}
	\label{l1}
	k_0\chi^2\sigma-4k_1\sigma(\sigma^2+\zeta^2+\delta^2)-2k_2(\sigma^3+3\sigma\delta^2)-2k_3\chi\sigma\zeta 
	-\frac{d}{3}\chi^4\Big(\frac{2\sigma}{\sigma^2-\delta^2}\Big)+\Big(\frac{\chi}{\chi_0}\Big)^2m_\pi^2f_\pi-\sum_i g_{\sigma i}\rho_i^s =0,
\end{multline}
\begin{multline}
	\label{l2}
	k_0\chi^2\zeta-4k_1\zeta(\sigma^2+\zeta^2+\delta^2)-4k_2\zeta^3-k_3\chi(\sigma^2-\delta^2)-\frac{d}{3}\frac{\chi^4}{\zeta}
	+\Big(\frac{\chi}{\chi_0}\Big)^2\Bigg[\sqrt{2}m_K^2f_K-\frac{1}{\sqrt{2}}m_\pi^2f_\pi\Bigg]-\sum_i g_{\zeta i}\rho_i^s=0,
\end{multline}
\begin{multline}
	\label{l3}
	k_0\chi^2\delta-4k_1\delta(\sigma^2+\zeta^2+\delta^2)-2k_2\delta(\delta^2+3\sigma^2) 
	+2k_3\chi\delta\zeta +\frac{2}{3}d\chi^4\Bigg(\frac{\delta}{\sigma^2-\delta^2}\Bigg)-\sum_i g_{\delta i}\rho_i^s=0,
\end{multline}
\begin{multline}
	\label{l4}
	k_0\chi(\sigma^2+\zeta^2+\delta^2)-k_3\zeta(\sigma^2-\delta^2)+\chi^3\Bigg[1+4\ln{\frac{\chi}{\chi_0}}\Bigg] 
	+(4k_4-d)\chi^3-\frac{4}{3}d\chi^3\ln{\Bigg[\Big(\frac{(\sigma^2-\delta^2)\zeta}{\sigma_0^2\zeta_0}\Big)\Big(\frac{\chi}{\chi_0}\Big)^3\Bigg]} \\
	+\Big(\frac{2\chi}{\chi_0^2}\Big)\Bigg[m_\pi^2f_\pi\sigma+\Bigg(\sqrt{2}m_K^2f_K-\frac{1}{\sqrt{2}}m_\pi^2f_\pi\Bigg)\zeta\Bigg] = 0,
\end{multline}
where $\rho_i^s\equiv\langle \bar{\psi_{i}} \psi_{i}\rangle$ is the scalar density of the $i^{th}$ baryon while  $\rho_i\equiv\langle \psi_{i}^\dagger\psi_{i}\rangle$ is the baryon density.

 Our primary focus of study -- the neutron star environment -- has a temperature of about 1 MeV at the core. This temperature however is very small compared to the chemical potential at the core ($\mu/T \gg 1$) and can be considered as a system of degenerate matter at zero temperature\footnote{We often refer to this system as the zero temperature regime in the present work.}. We can obtain analytical expressions for baryon density and scalar density at $T\xrightarrow{}0$ as,
\begin{eqnarray}
	&&\rho_i =2\frac{4\pi}{(2\pi)^3}\int_0^{p_{Fi}}p^2 dp = \frac{p_{Fi}^3}{3\pi^2}\label{de},\\
	&&\rho_i^s = 2\frac{4\pi}{(2\pi)^3}\int_0^{p_{Fi}}\frac{m_i^*}{E_i^*}p^2 dp = \frac{m_i^*}{2\pi^2}\Bigg[p_{Fi}E_{Fi}^*-m_i^{*2}\ln\Big(\frac{p_{Fi}+E_{Fi}^*}{m_i^*}\Big)\Bigg],\label{sde}
\end{eqnarray}
with $E^{*}_{Fi}=\sqrt{p_{Fi}^2+m_i^{*2}}$. Solving the coupled Eqs.~\eqref{l1}-\eqref{l4} along with Eqs.~\eqref{de} and~\eqref{sde}, one can obtain the values of $\sigma, \zeta, \delta$ and $\chi$ as a function of baryon density $\rho_{i}$. In the present calculation, we use $\delta$=0 for simplicity. So the Eqs.~\eqref{uu_bar} and~\eqref{dd_bar} modify to,
\begin{equation}\label{m_q}
	m_q\langle \bar{q}q \rangle = \frac{1}{2}m_\pi^2f_\pi\sigma,
\end{equation}
where $m_q=(m_u+m_d)/2$ is the current quark mass and $\langle \bar{q}q \rangle$ is the quark condensate. Using the density dependent $\sigma$ in Eq.~\eqref{m_q}, one can obtain the density-dependent quark condensate $\langle \bar{q}q\rangle$. Motivated by NJL-type models~\cite{Buballa:2003qv,Buballa:2014tba} earlier studies ~\cite{Dey:2019vkn} have defined the constituent quark mass from quark condensate calculated in LQCD simulations. For instance, the NJL-gap equation  $\langle \bar{q}q\rangle=-\frac{M-m}{2G}$ ~\cite{Buballa:2003qv}, can be simplified to $M\propto\langle \bar{q}q\rangle$ in the limit \(m\rightarrow 0\). Guided by this idea Ref.~\cite{Dey:2019vkn} has adopted a rough normalized relation for the constituent quark mass by normalizing it with the value of condensate at $T\rightarrow 0$. We have employed a similar approach here. By normalizing the density-dependent quark condensate with its vacuum values, we express the constituent $u$-quark mass as,
\begin{equation}\label{M_Q}
M_q(\rho_B)=\frac{\langle \bar{q}q\rangle(\rho_B)}{\langle \bar{q}q\rangle(\rho_B=0)}M_q(\rho_B=0),
\end{equation}
where $\rho_B$ is the baryon density, $[\langle \bar{q}q\rangle(\rho_{B}=0)]^{1/3}$ = 263.5 MeV and  $M_{q}(\rho_{B}=0)=M_{p}(\rho_{B}=0)/3$ = 313 MeV are the vacuum expectation values of the quark condensate and the constituent quark mass respectively. The constituent quark mass in vacuum is taken to be one-third of the nucleon mass.

We assume the quark density $\rho_q$ to be related to the baryon density $\rho_{B}$ as $\rho_q=3\rho_{B}$. 
The Fermi momentum of the quark matter is related to number density as $p_F=(6\pi^2 \rho_q /g)^{1/3}$, where $g=g_s\times g_c=2\times 3=6$ is the product of spin and colour degeneracy. This gives the expression for quark chemical potential, 
\begin{equation}\label{mu_Q}
	\mu_q=\sqrt{(3\pi^2\rho_{B})^{2/3}+M_{q}^{2}(\rho_{B})},
\end{equation}
where $M_q$ is the constituent quark mass obtained from the quark condensate. Eq.~\eqref{mu_Q} can be used to get both the ordinary chemical potential $\mu_{q}=\mu_{q}(\rho_{B},m_{q})$ and in-medium chemical potential $\mu_{q}= \mu_{q}(\rho_{B},M_{q}(\rho_{B}))$.

Now, we describe the changes required to obtain the quark condensates and constituent quark mass at finite temperature ($T\sim 100$ MeV) and chemical potential useful for system produced in CBM/NICA. The Eqs.~(\ref{de}) and~(\ref{sde}) become,
\begin{align}
\rho_i &=2\frac{4\pi}{(2\pi)^3}\int_0^{\infty}p^2~ (f^{*}_{i}-\bar{f}^{*}_{i}) dp\label{deprime},\\
\rho_i^s &=2\frac{4\pi}{(2\pi)^3}\int_0^{\infty}\frac{m_i^*}{E_i^*} (f^{*}_{i}-\bar{f}^{*}_{i})~p^2 dp,\label{sdeprime}
\end{align}
where we have the proton (neutron) distribution $f^{*}_{i}=1/[e^{(E^{*}_{i}-\mu_{i})/T}+1]$ and anti-proton (anti-neutron) distribution $\bar{f}^{*}_{i}=1/[e^{(E^{*}_{i}+\mu_{i})/T}+1]$ with $E_{i}^{*}=\sqrt{p^{2}+m_{i}^{*2}}$. The simultaneous solution of the coupled Eqs.~\eqref{l1}-\eqref{l4} along with Eqs.~\eqref{deprime} and~\eqref{sdeprime} provide us with the values of $\sigma, \zeta, \delta$ and $\chi$ as a function of baryon chemical potential $\rho_{B}$ and temperature $T$. The value of the quark condensate $\langle \bar{q}q \rangle(\rho_{B},T)$ is then obtained with the help of Eq.~(\ref{m_q}). The constituent quark mass at finite $T$ and $\mu_{B}$ is obtained by, 
\begin{align}
M_q(\rho_B,T)=\frac{\langle \bar{q}q\rangle(\rho_B,T)}{\langle \bar{q}q\rangle(\rho_B=0,T=0)}M_q(\rho_B=0,T=0)\label{M_Qprime}~.
\end{align} 
The expression for the quark chemical potential can be obtained self-consistently from,
\begin{align}
\rho_{q}=3\rho_{B}=g\frac{4\pi}{(2\pi)^{3}}\int p^{2}(f^{0}_{q}-\bar{f}^{0}_{q})dp\label{mu_Qprime}~,
\end{align}
where $f^{0}_{q}=1/[e^{(E-\mu_{q})/T}+1]$ and $\bar{f}^{0}_{q}=1/[e^{(E+\mu_{q})/T}+1]$ stands for quark and anti-quark distributions respectively with $E=\sqrt{p^{2}+M_{q}^{2}(\rho_{B},T)}$.

In the next section, we will use this constituent quark mass to determine the electrical conductivity components within a kinetic theory framework.

\subsection{Electrical conductivity in relaxation time approximation (RTA)}\label{IIB}
The  electrical current density for a system of quarks, in the microscopic kinetic theory, is defined as~\cite{Dey:2021fbo}
\begin{equation}
	J^{i}=gQ\int  \frac{d^3\vec{p}}{(2\pi)^{3}} \frac{p^{i}}{E} \delta f,\label{as0}
\end{equation} 
where $\delta f$ is the out-of-equilibrium contribution to the local equilibrium distribution function of the quark with degeneracy $g$, momentum $\vec{p}$ and energy $E=\sqrt{\vec{p}^{2}+M_{q}^{2}}$. $M_{q}$ is the mass of the constituent quark at finite baryon density, which is significantly higher than the current quark mass $m$. Here, we consider a single-flavoured system consisting of $u$-quarks with $Q=2e/3$ (and its anti-particle -- $\bar{u}$-quarks). Eq.~\eqref{as0} forms the basis of microscopic calculation of $J^{i}$ in the Boltzmann Transport Equation (BTE)~\cite{kremer2010int} based kinetic theory. This microscopic form, when compared with the macroscopic version of the current density given by Ohm's law, i.e., $J^{i}=\sigma^{ij} \tilde{E}^{j}$ with $\tilde{E}^{j}$ being the electric field, provides us with the electrical conductivity tensor $\sigma^{ij}$. The kinetic evaluation of $\delta f$ and the subsequent determination $\sigma^{ij}$ can be carried out by solving the BTE for the quarks~\cite{Dey:2021fbo},
\begin{eqnarray}
	&&-\frac{f^{0}(1-f^{0})}{ET}Q\tilde{E}^{j}p^{j}+ Q\epsilon^{ijk} v^{j} B^{k} \frac{\partial \delta f}{\partial p^{i}}=-\frac{\delta f}{\tau_{c}},\label{as1}
\end{eqnarray}
where $v^{j}=p^{j}/E$ and $f^{0}=1/(e^{(E-\mu_q)/T}+1)$. Similar to Sec.~(\ref{IIA}) we first consider the degenerate quark matter resembling the neutron star environment in the calculation of conductivity. In this low temperature limit conductivity takes analytical form which we provide in Secs.~(\ref{concl}) and (\ref{conqm}). Subsequently, in Sec.~(\ref{confiniteT}) we discuss the conductivity at finite chemical potential and temperature, corresponding to the matter produced in CBM/NICA.
At $T\xrightarrow[]{} 0$ the conductivity tensor has the form,
 \begin{eqnarray}
 	\sigma^{ij}&=&gQ^{2}\int \frac{d^{3}\vec{p}}{(2\pi)^{3}} \delta(\mu_q-E) \frac{\tau_{c}}{1+(\tau_{c}/\tau_{B})^{2}} \frac{p^ip^k}{E^{2}}\bigg[\delta^{jk}+ \left(\frac{\tau_{c}}{\tau_{B}}\right)^{2} b^{j}b^{k} +\frac{\tau_{c}}{\tau_{B}} \epsilon^{kjm}b^{m} \bigg],\label{as2}
 \end{eqnarray} 
which follows from $f^{0}\xrightarrow{T=0} \theta (\mu_q-E)$ and $\frac{\partial f^{0}}{\partial E} \xrightarrow{T=0} \frac{\partial \theta (\mu_q-E) }{\partial E}=-\delta (\mu_q-E)$. Here, $\tau_c$ is the thermal relaxation time and $\tau_B=E/QB$ is the magnetic relaxation time. Readers can look into Refs.~\cite{Ghosh:2019ubc, Dey:2021fbo, Dey:2019axu} for detailed discussion and derivation of the expression. 
 In the following two subsections, we use Eq.~\eqref{as2} to determine the classical and quantized expressions of electrical conductivity components. 
 \subsubsection{\textbf{Classical expressions of electrical conductivity at $T\rightarrow 0$}}\label{concl} 
For a relatively weak magnetic field $B$, one can obtain the classical components of the conductivity tensor $\sigma^{ij}$ by using the relation,
  \begin{eqnarray}
  \int \frac{d^{3}\vec{p}}{(2\pi)^{3}} p^{i}p^{k}=\int \frac{d^{3}p}{(2\pi)^{3}} \frac{p^{2}}{3} \delta^{ik}, \label{as3}
  \end{eqnarray}
  where $d^{3}p\equiv 4\pi^{2}pdp$, as,
 \begin{eqnarray}
 	\sigma^{ij}&=&gQ^{2}\int \frac{d^{3}p}{(2\pi)^{3}} \delta (\mu_q-E) \frac{\tau_{c}}{1+(\tau_{c}/\tau_{B})^{2}} \frac{p^2}{3E^2}\bigg[\delta^{ij}+ \left(\frac{\tau_{c}}{\tau_{B}}\right)^{2} b^{i}b^{j} +\frac{\tau_{c}}{\tau_{B}} \epsilon^{ijk}b^{k} \bigg].\label{as4}
 \end{eqnarray} 
  From Eq.~\eqref{as4}, we can
  obtain the parallel $\sigma^{||}_{CM}\equiv \sigma^{zz}$, perpendicular $\sigma^{\perp}_{CM}\equiv\sigma^{xx}=\sigma^{yy}$ and Hall $\sigma^{\times}_{CM}\equiv\sigma^{xy}=-\sigma^{yx}$ components of conductivity for a magnetic field $\vec{B}=B\hat{k}$ along the $z$-direction as,
\begin{eqnarray}
&&\sigma^{||}_{CM}= gQ^{2}\int \frac{d^{3}p}{(2\pi)^{3}} \tau_{c}\delta(\mu_q-E)\frac{p^2}{3E^2} =\frac{Q^{2}}{\pi^{2}}\tau_{c}\frac{(\mu_q^{2}-M_{q}^{2})^{3/2}}{\mu_q},\label{as5}\\
&&\sigma^{\perp}_{CM}=gQ^{2}\int \frac{d^{3}p}{(2\pi)^{3}} \frac{\tau_{c}}{1+(\tau_{c}/\tau_{B})^{2}}\delta(\mu_q-E)\frac{p^2}{3E^2}=\frac{Q^{2}}{\pi^{2}}\tau^{\perp}_{c}\frac{(\mu_q^{2}-M_{q}^{2})^{3/2}}{\mu_q},\label{as6}\\
&&\sigma^{\times}_{CM}=gQ^{2}\int \frac{d^{3}p}{(2\pi)^{3}} \delta(\mu_q-E) \frac{\tau_{c} (\tau_{c}/\tau_{B})}{1+(\tau_{c}/\tau_{B})^{2}} \frac{p^2}{3E^2}=\frac{Q^{2}}{\pi^{2}}\tau_{c}^{\times}\frac{(\mu_q^{2}-M_{q}^{2})^{3/2}}{\mu_q}\label{as7}, 
\end{eqnarray}
where 
$\tau_c$, $\tau^{\perp}_{c}\equiv \frac{\tau_{c}}{1+(\tau_{c}/\tau_{B})^{2}}$ and $\tau^{\times}_{c}\equiv \frac{\tau_{c} (\tau_{c}/\tau_{B})}{1+(\tau_{c}/\tau_{B})^{2}}$ are the effective relaxation times along the parallel, perpendicular and Hall directions respectively. 
The delta function replaces $E$ with the Fermi energy or the quark chemical potential $\mu_q$, i.e., at $T\rightarrow 0$, $\tau_{B}=\frac{E}{QB}\rightarrow \frac{\mu_q}{QB}$. It should be noted that a physical solution exists only for $\mu_q>M_{q}$. 
\subsubsection{\textbf{Quantized expressions of electrical conductivity at $T\rightarrow 0$}}\label{conqm}
In the presence of strong magnetic fields, the conductivity of the system can be altered significantly due to the quantized cyclotron orbits of the quarks in the plane perpendicular to the direction of the magnetic field. The conductivity tensor $\sigma^{ij}$ in Eq.~\eqref{as2} can be re-expressed by modifying the phase space integrals to incorporate the quantization of the quark momenta in the $xy$-plane while keeping $p^{z}$ as a continuous variable, \textit{i.e.,} 
\begin{eqnarray}
&&g_{s}\int \frac{d^3\vec{p}}{(2\pi)^{3}} \rightarrow \frac{QB}{(2\pi)^{2}}\sum_{l=0}^\infty \alpha_{l}\int d\phi \int \frac{dp^{z}}{2\pi}.\label{as8}
\end{eqnarray}
The spin degeneracy \( g_s \) changes to \( \alpha_l \), which is 1 for the lowest Landau level (LLL) \( l = 0 \), and 2 for all other Landau levels \( l \). In general, one can write  $\alpha_l = 2 - \delta_{l,0}.$ 	\( \phi \) is the azimuthal angle. Eq.~(\ref{as2}), along with the phase space discretization performed in Eq.~(\ref{as8}), takes the form,
 \begin{eqnarray}
 	\sigma^{ij}&=&g_{c}Q^{2}\int \frac{QB}{(2\pi)^{2}}\sum_{l=0}^\infty \alpha_{l}\int d\phi \int \frac{dp^{z}}{2\pi} \delta(\mu_q-E) \frac{\tau_{c}}{1+(\tau_{c}/\tau_{B})^{2}} \frac{p^{i}p^{k}}{E^{2}}\bigg[\delta^{jk}+ \left(\frac{\tau_{c}}{\tau_{B}}\right)^{2} b^{j}b^{k} +\frac{\tau_{c}}{\tau_{B}} \epsilon^{kjm}b^{m} \bigg].\label{as9}
 \end{eqnarray} 
 The quantized expressions of the parallel $\sigma_{QM}^{||}\equiv \sigma^{zz}$, perpendicular $\sigma_{QM}^{\perp}\equiv\sigma^{xx}=\sigma^{yy}$ and Hall $\sigma_{QM}^{\times}\equiv\sigma^{xy}$  components of conductivity when $\vec{B}=B\hat{k}$ can be obtained from Eq.~\eqref{as9} as,
 \begin{eqnarray}
&&\sigma_{QM}^{||}=3Q^{2}\sum_{l=0}^\infty \alpha_{l}\frac{QB}{(2\pi)^{2}}\int_{-\infty}^{\infty} \delta(\mu_q-E_l)\tau_{c}\frac{(p^{z})^2}{E_l^2}dp^{z}=6Q^{2}\sum_{l=0}^{l_{max}}\alpha_{l}\frac{QB}{(2\pi)^{2}}\tau_{c}\frac{\sqrt{\mu_{q}^{2}-M_{l}^{2}}}{\mu_{q}},\label{as10}\\
&&\sigma_{QM}^{\perp}=3Q^{2}\sum_{l=0}^\infty l\alpha_{l} \frac{(QB)^{2}}{(2\pi)^{2}} \int_{-\infty}^{\infty} \delta(\mu_q-E_l)\frac{\tau_{c}}{1+(\tau_{c}/\tau_{B})^{2}}\frac{dp^{z}}{E^{2}_l}=6Q^{2}\sum_{l=0}^{l_{max}}l\alpha_{l}\frac{Q^2B^2}{(2\pi)^{2}}\tau_{c}^{\perp}\frac{1}{\mu_q\sqrt{\mu_q^{2}-M_{l}^{2}}},\label{as11}\\
&&\sigma_{QM}^{\times}=3Q^{2}\sum_{l=0}^\infty l\alpha_{l} \frac{(QB)^{2}}{(2\pi)^{2}} \int_{-\infty}^{\infty} \delta(\mu_q-E_l)\frac{\tau_{c}(\tau_{c}/\tau_{B})}{1+(\tau_{c}/\tau_{B})^{2}}\frac{dp^{z}_{r}}{E^{2}_l}=6Q^{2}\sum_{l=0}^{l_{max}}l\alpha_{l}\frac{Q^2B^2}{(2\pi)^{2}}\tau_{c}^{\times}\frac{1}{\mu_q\sqrt{\mu_q^{2}-M_{l}^{2}}},\label{as12}
\end{eqnarray}
 where we have $B$-dependent effective mass $M_{l}\equiv \sqrt{M_{q}^{2}+2lQB}$, Landau level energies $E_l\equiv\sqrt{(p^{z})^{2}+M_{l}^{2}}$ and effective relaxation times, $\tau^{\perp}_{c}\equiv \frac{\tau_{c}}{1+(\tau_{c}/\tau_{B})^{2}}$ and $\tau^{\times}_{c}\equiv \frac{\tau_{c} (\tau_{c}/\tau_{B})}{1+(\tau_{c}/\tau_{B})^{2}}$. A physical solution exists only when $\mu_{q}>M_{l}$, which puts a constraint on the maximum number of Landau levels included, 
 \begin{eqnarray}
 && M^{2}_{l}=M_{q}^{2}+2lQB < \mu_q^{2}\nonumber\\
 \implies && l < \frac{\mu_q^{2}-M_{q}^{2}}{2QB}\nonumber\\
  \implies && l_{max}= \text{Integer} \left[\frac{\mu_q^{2}-M_{q}^{2}}{2QB}\right].\label{as13}
 \end{eqnarray}
 The conductivity matrix $[\sigma]$ for both classical and quantized expressions can be written as:
 \begin{equation}
 	[\sigma]=\begin{pmatrix}
 		\sigma^{\perp} & \sigma^{\times} & 0\\
 		-\sigma^{\times} & \sigma^{\perp} & 0\\
 		0 & 0 & \sigma^{||}
 	\end{pmatrix}~.
 	\label{as14}
 \end{equation}
Another critical quantity, the resistivity, is defined by the relation $\tilde{E}^{i} \equiv \rho^{ij} J^{j}$. When no magnetic field is present, the conductivity is isotropic, and the resistivity is simply the inverse of the isotropic conductivity. In general, the resistivity matrix \([\rho]\) can be obtained by inverting the conductivity matrix \([\sigma]\) as:

\begin{equation}
[\rho] =\begin{pmatrix}
\frac{\sigma^{\perp}}{(\sigma^\perp)^2+(\sigma^\times)^2} & -\frac{\sigma^{\times}}{(\sigma^\perp)^2+(\sigma^\times)^2} & 0\\
\frac{\sigma^{\times}}{(\sigma^\perp)^2+(\sigma^\times)^2} & \frac{\sigma^{\perp}}{(\sigma^\perp)^2+(\sigma^\times)^2} & 0\\
0 & 0 & \frac{1}{\sigma^{||}}
\end{pmatrix}.\label{as14}
\end{equation}
The three components of resistivity has the form given by,
\begin{align}
	\rho^{||}&=\rho^{zz}=\frac{1}{\sigma^{||}},\label{as15}\\
	\rho^{\perp}&=\rho^{xx}=\rho^{yy}=\frac{1}{1+\left(\frac{\tau_c}{\tau_B}\right)^2}\frac{1}{\sigma^{\perp}}=\frac{\tau_{c}^{\perp}}{\tau_{c}}\frac{1}{\sigma^{\perp}},\label{as16}\\
	\rho^{\times}&=-\rho^{xy}=\rho^{yx}=\frac{\frac{\tau_c}{\tau_B}}{1+\left(\frac{\tau_c}{\tau_B}\right)^2}\frac{1}{\sigma^{\perp}}=\frac{\tau_{c}^{\times}}{\tau_{c}}\frac{1}{\sigma^{\perp}}\label{as17}.
\end{align}
 Eqs.~\eqref{as15} - \eqref{as17}, along with Eqs.~\eqref{as5} - \eqref{as7} and Eqs.~\eqref{as10} - \eqref{as12} help us express the classical and quantized resistivity matrix elements as follows:
\begin{eqnarray}
&& \rho_{CM}^{||}=\rho_{CM}^{\perp}=\frac{\pi^{2}}{Q^{2}\tau_{c}} \frac{\mu_q}{(\mu_q^{2}-M_{q}^{2})^{3/2} }\label{as18},\\
&&\rho_{CM}^{\times}=\frac{\pi^{2}}{Q}\frac{B}{(\mu_q^{2}-M_{q}^{2})^{3/2}}\label{as19},\\
&& \rho_{QM}^{||}=\frac{2\pi^{2}}{3 Q^3B\tau_{c}}\left(\sum_{l=0}^{l_{max}}\alpha_{l}\frac{\sqrt{\mu_q^{2}-M_{l}^{2}}}{\mu_q}\right)^{-1}\label{as20},\\
&& \rho_{QM}^{\perp}=\frac{2\pi^{2}}{3 Q^{4}B^{2}\tau_{c}}\left(\sum_{l=0}^{l_{max}}l\alpha_{l}\frac{1}{\mu_q{\sqrt{\mu_q^{2}-M_{l}^{2}}}}\right)^{-1}\label{as21},\\
&& \rho_{QM}^{\times}=\frac{2\pi^{2}}{3 Q^{3}B}\left(\sum_{l=0}^{l_{max}}l\alpha_{l}\frac{1}{{\sqrt{\mu_q^{2}-M_{l}^{2}}}}\right)^{-1}.\label{as22}
\end{eqnarray}

\subsubsection{Classical and quantized expressions of electrical conductivity for finite $\mu_{q}$ and $T$}\label{confiniteT}
At finite temperature the Fermi-Dirac distribution function will significantly vary from the step function used in the previuos conductivity calculation; therefore, replacing $\theta(\mu_{q}-E)\xrightarrow{}f^{0}$, $\delta(\mu_{q}-E)\xrightarrow{}-\frac{\partial f^0}{\partial E}$ and adding the contributions from anti-quarks, the classical expressions~(\ref{as5})--(\ref{as7}) of conductivity become
\begin{align}
&\sigma^{||}_{CM}= \frac{g Q^{2}}{T}\int \frac{d^{3}p}{(2\pi)^{3}} \tau_{c}\frac{p^2}{3E^2}~[f_{q}^{0}(E)(1-f_{q}^{0}(E))+\bar{f}_{q}^{0}(E)(1-\bar{f}_{q}^{0}(E))]\label{as5prime}~,\\
& \sigma^{\perp}_{CM}= \frac{g Q^{2}}{T}\int \frac{d^{3}p}{(2\pi)^{3}} \frac{\tau_{c}}{1+(\tau_{c}/\tau_{B})^{2}}\frac{p^2}{3E^2}~[f_{q}^{0}(E)(1-f_{q}^{0}(E))+\bar{f}_{q}^{0}(E)(1-\bar{f}_{q}^{0}(E))]\label{as6prime}~,\\
& \sigma^{\times}_{CM}= \frac{g Q^{2}}{T}\int \frac{d^{3}p}{(2\pi)^{3}} \frac{\tau_{c}(\tau_{c}/\tau_{B})}{1+(\tau_{c}/\tau_{B})^{2}}\frac{p^2}{3E^2}~[f_{q}^{0}(E)(1-f_{q}^{0}(E))-\bar{f}_{q}^{0}(E)(1-\bar{f}_{q}^{0}(E))]\label{as7prime}~,
\end{align}
where we have suppressed the energy dependence of $\tau_{B}=E/QB$. In Eq.~(\ref{as7prime}) we have a negative sign for the anti-particle contributions to the hall conductivity. This arises because of the directionality involved in the cyclotron time-period $\tau_{B}$ which changes sign according to the sign of electric charge. Similarly, the quantized expressions~(\ref{as10})--(\ref{as12}) of the conductivity become 
\begin{align}
&\sigma_{QM}^{||}=\frac{3 Q^{2}}{T}\sum_{l=0}^\infty \alpha_{l}\frac{|Q|B}{(2\pi)^{2}}\int_{-\infty}^{\infty} [f_{q}^{0}(E_{l})(1-f_{q}^{0}(E_{l}))+\bar{f}_{q}^{0}(E_{l})(1-\bar{f}_{q}^{0}(E_{l}))]\tau_{c}\frac{(p^{z})^2}{E_l^2}dp^{z},\label{as10prime}\\
&\sigma_{QM}^{\perp}=\frac{3 Q^{2}}{T}\sum_{l=0}^\infty l\alpha_{l} \frac{(|Q|B)^{2}}{(2\pi)^{2}} \int_{-\infty}^{\infty} [f_{q}^{0}(E_{l})(1-f_{q}^{0}(E_{l}))+\bar{f}_{q}^{0}(E_{l})(1-\bar{f}_{q}^{0}(E_{l}))]\frac{\tau_{c}}{1+(\tau_{c}/\tau_{B,l})^{2}}\frac{dp^{z}}{E^{2}_l},\label{as11prime}\\
&\sigma_{QM}^{\times}=\frac{3 Q^{2}}{T}\sum_{l=0}^\infty l\alpha_{l} \frac{(|Q|B)^{2}}{(2\pi)^{2}} \int_{-\infty}^{\infty} [f_{q}^{0}(E_{l})(1-f_{q}^{0}(E_{l}))-\bar{f}_{q}^{0}(E_{l})(1-\bar{f}_{q}^{0}(E_{l}))]\frac{\tau_{c}(\tau_{c}/\tau_{B,l})}{1+(\tau_{c}/\tau_{B,l})^{2}}\frac{dp^{z}}{E^{2}_l},\label{as12prime}
\end{align}
where we have $\tau_{B,l}\equiv E_{l}/QB$. In the summing over the landau levels in  Eqs.~(\ref{as10prime})--(\ref{as12prime}) there is no restriction on the number of the levels involved. This should be contrasted with the $T \to 0$ case, where the magnitude of $\mu_q$ and $B$ determines the maximum number of occupied Landau levels.

\section{Results and Discussions}\label{III}

In this section, we provide a quantitative comparison of the conductivity components obtained from the relaxation time approximated (RTA) BTE in two settings -- 1. at $T=1$ MeV, and 2. at $T=100$ MeV. The estimated values of the conductivities are relevant for the dense cold matter expected in the core of neutron stars (1st setting) and the reaction zone of accelerator facilities at CBM/NICA (2nd setting). Given that the thermal (Fermi-Dirac) distribution function in the first setting closely resembles a step function, as in the $T=0$ limit, we interpret the results accordingly for simplicity.

It is worth mentioning that kinetic theory frameworks are more economical than the tedious diagrammatic Kubo approach~\cite{Jeon:1992kk}. The study of transport coefficients via Kubo formalism in absence~\cite{Ghosh:2014yea,Ghosh:2018xll,Ghosh:2014ija,Ghosh:2015mda,Ghosh:2016yvt,Singha:2017jmq} and presence~\cite{Ghosh:2020wqx, Satapathy:2021cjp,Satapathy:2021wex,Bandyopadhyay:2023lvk} of magnetic fields have been done in Refs.~\cite{Ghosh:2014yea,Ghosh:2018xll,Ghosh:2014ija,Ghosh:2015mda,Ghosh:2016yvt,Singha:2017jmq,Ghosh:2020wqx, Satapathy:2021cjp,Satapathy:2021wex,Bandyopadhyay:2023lvk}. Instead of deriving the electrical conductivity tensor from the Lagrangian density via the Kubo approach, an alternative approach is to construct a quasiparticle description—using constituent quark or hadron masses obtained from the Lagrangian density~\cite{Chakraborty:2010fr,Rocha:2022fqz}— within the Relaxation Time Approximation (RTA) framework for the conductivity tensor. This is the approach we have adopted in this work.
We consider a simplified system 
consisting of a single quark flavour - the $u$ quark and compare the conductivity components in two different scenarios, one where we use the  current quark mass $m_{q}=(m_{u}+m_{d})/2=4.63$ MeV  and  the other where we consider density-dependent (or, density and temperature-dependent) in-medium constituent quark mass $M_{q}(\rho_{B})$ (or, $M_q(\rho_B,T)$) obtained using the chiral effective model described in Sec.~\ref{IIA}. The results are compared in the classical (unquantized) and quantized picture. We observe the variation in conductivity components as a function of magnetic field and baryon density. 

At $T=1$ MeV, the quark number density can be approximated as  $\rho_q=g\int_0^\infty \frac{d^3p}{(2\pi^2)}\Theta(\mu_q-E)=\frac{1}{\pi^2}(\mu_q^2-M_q^2)^{3/2}$ where $\Theta(\mu_q-E)$ is the equilibrium distribution function for the quarks at $T=0$. A physical solution exists only when $\mu_q>M_{q}$. However, in presence of a magnetic field, the perpendicular momenta are quantized leading to quantized number density as discussed earlier. The quantized version of number density at $T=0$ is given by $\rho_q= g_c\sum_{0}^{\infty}\alpha_l\frac{qB}{(2\pi)^2}\int_0^\infty dk_z\Theta(\mu_q-E_l)=3\frac{qB}{2\pi^2}\sum_{0}^{\infty}\alpha_l\sqrt{\mu_q^2-2lqB-M_q^2}$. A physical solution exists only for $\mu_q^2>2lqB+M_q^2$ which puts a constraint on the maximum number of Landau levels included $l_{max}= \text{Integer}\left[\frac{\mu_q^2-M_q^2}{2qB}\right]$ \cite{Dey:2021fbo}. However, we have used the baryon density in the absence of a magnetic field as the common x-axis for simplicity. Similarly, the number density at finite temperature is given by Eq. \eqref{mu_Qprime}. 

\begin{figure}
	\centering
	\includegraphics[width=0.45\linewidth]{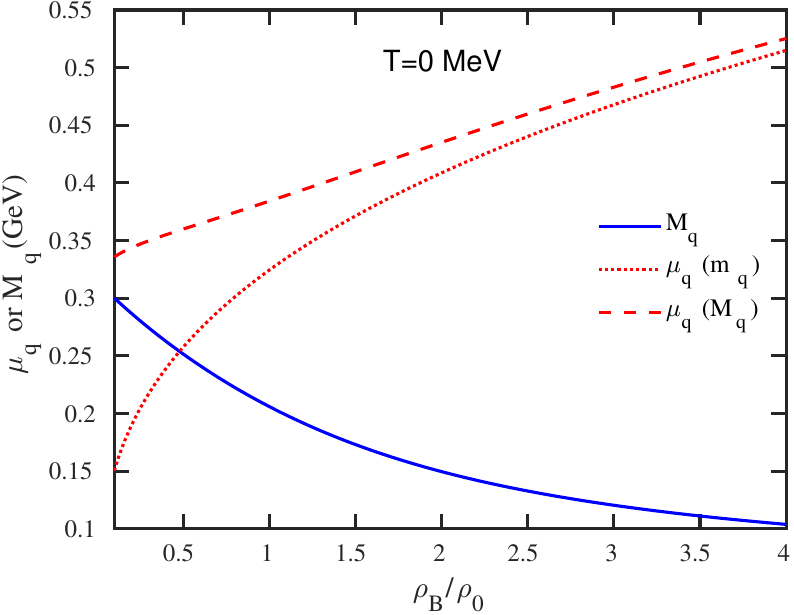}
	\includegraphics[width=0.45\linewidth]{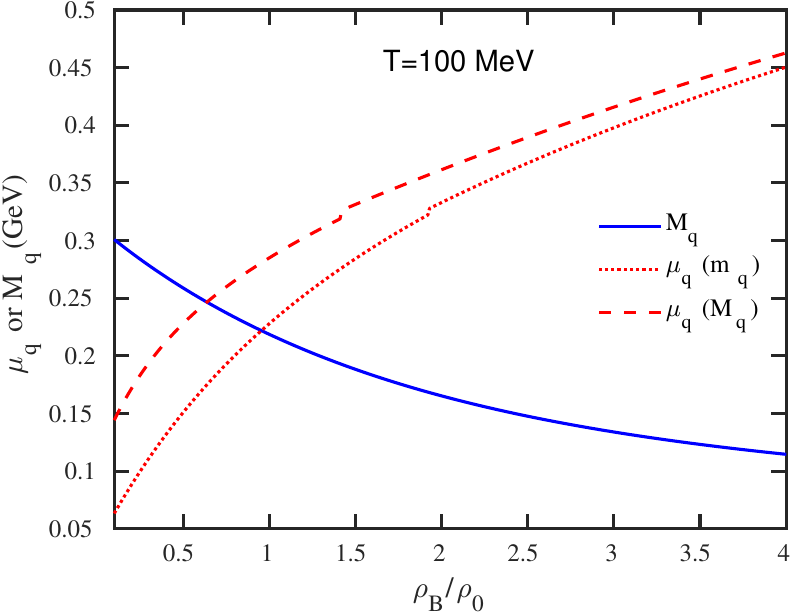}
	\caption{The variation of the quark chemical potential $\mu_{q}$ and constituent quark mass $M_{q}$ as a function of the scaled baryon density $\rho_{B}/\rho_{0}$ for quark matter at $T=0$ (left panel) and $T=100$ MeV (right panel) obtained using the chiral effective model.}\label{A25}
\end{figure}

Fig.~(\ref{A25}), displays the variation of current quark chemical potential $\mu_{q}(\rho_{B},m_{q},T)$ (red dotted line), constituent quark mass $M_{q}=M_{q}(\rho_{B},T)$ (blue solid line) and constituent quark chemical potential $\mu_{q}=\mu_{q}(\rho_{B}, M_{q}(\rho_{B}),T)$ (red dashed line) as a function of the scaled baryon density $\rho_{B}/\rho_{0}$, where $\rho_{0}=0.15$~fm$^{-3}$ is the nuclear saturation density at temperatures $T= 0$ (left panel) and $T=100$ MeV (right panel). At $T=0$, the current quark chemical potential $\mu_{q}(\rho_{B},m_q)$ increases monotonically with a gross dependence of $\rho_{B}^{1/3}$ at high $\rho_{B}$. This behaviour follows from Eq.~\eqref{mu_Q}, where one gets $\mu_{q}=\rho_{B}^{1/3}$ in the limit of zero current quark mass $m_{q}$. The constituent quark mass $M_{q}$ is proportional to the quark condensate $\langle\bar{q}q\rangle$. As the $\langle\bar{q}q\rangle$ value decreases with increasing baryon density, the plot shows a decrease in $M_{q}(\rho_{B})$ with increasing $\rho_{B}$ following Eq.~\eqref{M_Q}.
For the in-medium constituent quark chemical potential $\mu_{q}(\rho_{B}, M_{q}(\rho_{B}))$ (red dashed line), we notice that it has a positive vertical shift in comparison to the $\mu_{q}(\rho_{B},m_{q})$ (red dotted line) accounting for the large constituent quark mass $M_{q}=312.97$ MeV at $\rho_{B}\rightarrow 0$ in comparison to the current quark mass $m_{q}=4.63$ MeV. The difference between the two chemical potentials decreases as one goes to higher baryon density owing to the decrease in $M_{q}$ with increasing $\rho_{B}$. This feature is evident in the plot where the red dotted curve approaches the red dashed curve as $\rho_{B}$ increases.  
The constituent quark mass shows negligible variation as we move from $T=1$ MeV to $T=100$ MeV.  The effects of temperature on the value of scalar fields and thus the condensates are marginal till about 110-120 MeV, and more pronounced thereafter. Their magnitudes remain nearly unchanged from vacuum values up to about 100 MeV \cite{Kumar:2010gb}. However the constraint on the chemical potential $\mu_q>M_{q}$ is lifted at finite temperature and the current and constituent chemical potentials $\mu_{q}(\rho_{B}, m_q ~\text{or}~ M_{q}(\rho_{B}),T)$ -- obtained self-consistently from Eq.~\eqref{mu_Qprime} -- can be smaller than the corresponding masses $m_q~ \text{or}~ M_q(\rho_B)$.

We now proceed to examine how electrical conductivity and resistivity vary with baryon density and magnetic field in the context of neutron star matter. Even when the neutron star has a non-zero absolute temperature ($T= 1$ MeV) since $\mu_{q}/T \gg 1$, the distribution function can be replaced by a step function and the assumption of degenerate quark matter can give coherent results in the regime of neutron star densities. Similar calculations were previously reported in \cite{Na:2012td,Sen:2021tdu,Liu:2023kzy,Dey:2021fbo}. Later, we also present the variation of electrical conductivity at finite temperature ($\sim 100$ MeV), relevant to the dense nuclear matter expected in CBM and NICA experiments.

It is well established that at exactly zero temperature the relaxation time diverges in the Fermi liquid theory. However, gauge theories such as QCD  show non-Fermi liquid effects in degenerate quark plasmas \cite{Heiselberg:1993cr} and in the normal phase of high density QCD \cite{Schafer:2004zf}. The relaxation time in such scenarios differs significantly from the power law as given by the Fermi liquid theory and diverges much slowly as one goes towards zero temperature. For a system of degenerate quark matter, the time scale for momentum relaxation is $\tau_c =0.3\left(\frac{\mu_q}{300 ~\text{MeV}}\right)^{2/3}\left(\frac{\alpha_s T}{100 ~\text{MeV}}\right)^{-5/3}$ fm/c when $T\ll q_D\ll\mu_q$, where $\alpha_s$ is the strong coupling constant, $T$ is the temperature, $\mu_q$ is quark chemical potential and  $q_D$ is the Debye wave number \cite{Heiselberg:1993cr}. For a neutron star environment where $\alpha_s$ = 1 and $T$ = 1 MeV, \cite{Heiselberg:1993cr, Alford:2014doa} the momentum relaxation time $\tau_s$ ranges from approximately 300 fm to 1000 fm as the quark chemical potential increases from 300 MeV to 600 MeV. In our present calculation we use an average value of relaxation time $\tau_c$ = 500 fm over the entire density range. In the CBM/NICA environment where $\alpha_s=0.5$ and $T=100$ MeV, we use an average momentum relaxation time of 1 fm. The electric charge is related to the coupling constant as $e^2 = 4\pi/137$.
\begin{figure*}
	\centering
	\includegraphics[width=0.85\textwidth]{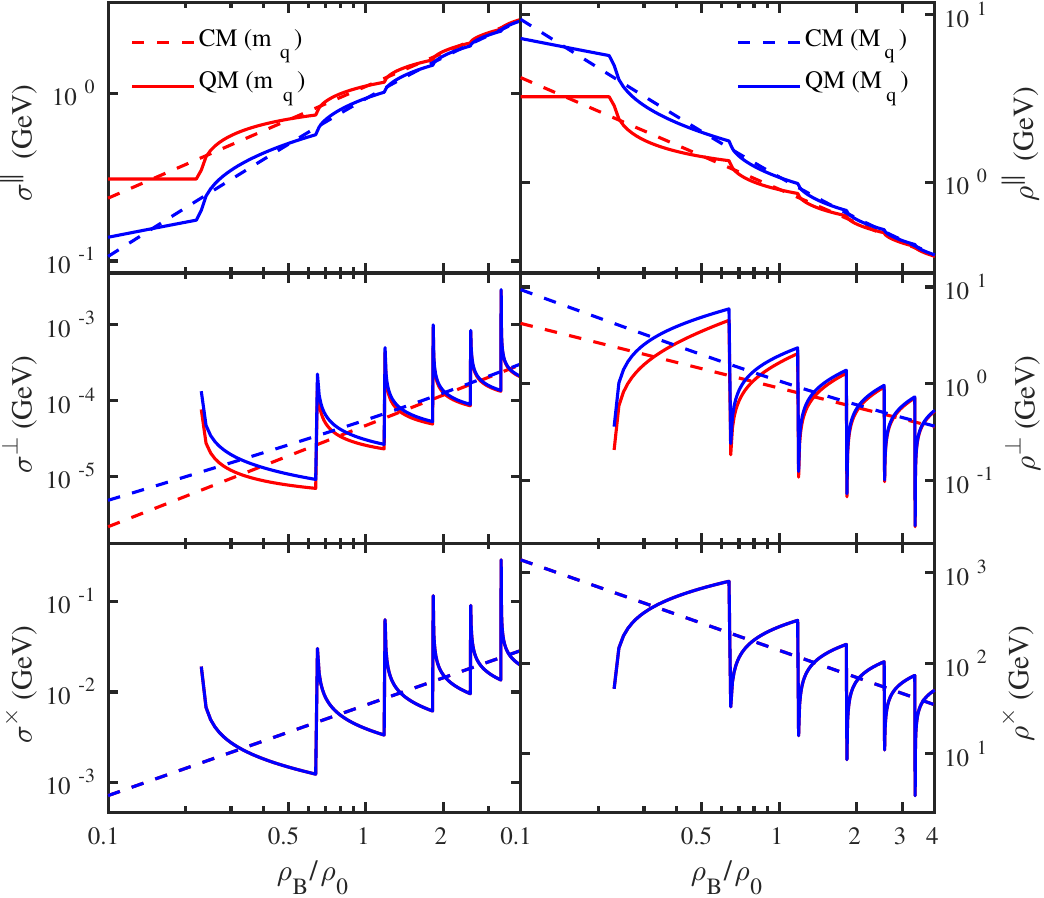}
	\caption{The variation of different components of electrical conductivity (left panel) and resistivity (right panel) as a function of scaled baryon density $\rho_{B}/\rho_{0}$ for a constant magnetic field $eB=5~m_{\pi}^{2}$.}\label{sigma_rho}
\end{figure*}

Magnetic fields create anisotropy in a system by singling out a specific direction in space, making the conductivity components different in different directions. For a magnetic field along the $z$-axis, we have three different conductivity components --- parallel conductivity $\sigma^{||}=\sigma^{zz}$, perpendicular conductivity $\sigma^{\perp}=\sigma^{xx}=\sigma^{yy}$ and Hall conductivity $\sigma^{\times}=\sigma^{xy}=-\sigma^{yx}$ --- whose expressions are set down in Sec.~\ref{IIB}. For the classical consideration, 
Eqs.~\eqref{as5}-\eqref{as7} can be used to see the variation of conductivity in two different scenarios --- first, by using the current quark mass $m_{q}$ and chemical potential $\mu_{q}$, i.e., $\sigma^{||,\perp,\times}_{CM}(\mu_{q}(\rho_{B},m_{q}),m_{q},B)\equiv \sigma^{||,\perp,\times}_{CM}(\rho_{B},m_{q},B)$ and second, by employing the 
constituent quark mass $M_{q}$ and chemical potential $\mu_{q}$ obtained with the help of chiral model in Sec.~\ref{IIA} i.e., $\sigma^{||,\perp,\times}_{CM}(\mu_{q}(\rho_{B},M_{q}(\rho_{B})),M_{q}(\rho_{B}),B)\equiv \sigma^{||,\perp,\times}_{CM}(\rho_{B},M_{q},B)$. Similarly, for the quantized picture of the magnetic field Eqs.~\eqref{as10}-\eqref{as12} can be used to see the variation in the conductivities $\sigma_{QM}^{||,\perp,\times}(\rho_{B},m_{q},B)$ and $\sigma_{QM}^{||,\perp,\times}(\rho_{B},M_{q},B)$ for the current quark and constituent quark descriptions respectively. Throughout this discussion, the subscripts $CM$ and $QM$ stand for the classical expression (without Landau quantization) and the quantized description (including Landau quantization) respectively. 

In Fig.~(\ref{sigma_rho}), we have displayed both the conductivity (left panel) and resistivity (right panel) components as a function of scaled baryon density $\rho_{B}/\rho_{0}$ at a fixed magnetic field $eB=5~m^{2}_{\pi}$. The changes in the classical parallel conductivity $\sigma^{||}_{CM}=\sigma^{||}_{CM}(\rho_{B}, m_{q})$ (red dashed line) with respect to $\rho_{B}/\rho_{0}$ is shown at the top of the left panel. In the high baryon density limit, an approximate variation $\sigma_{CM}^{||}\propto \rho_B^{2/3}$ is observed as $\sigma^{||}_{CM}\propto \rho_{B}/\sqrt{(3\pi^{2}\rho_{B})^{2/3}+m_{q}^{2}}\xrightarrow{m_{q}=0} \left(\frac{1}{3\pi^{2}}\right)^{1/3}\rho_{B}^{2/3}$. 
$\sigma^{||}_{CM}(\rho_{B},M_{q})$ (blue dashed line) calculated with the effective mass obtained from the chiral model follows the same overall trend. However, the magnitude is significantly suppressed in the low baryon density regime ($\rho_{B}=0.1\rho_{0}-1 \rho_{0}$), and the conductivity approaches the curve for $\sigma^{||}(\rho_{B},m_{q})$ as $\rho_{B}$ increases. Similar suppressions are also seen in $\sigma^{\perp,\times}(\rho_{B},M_{q})$ in comparison to $\sigma^{\perp,\times}(\rho_{B},m_{q})$. The increasing values of the classical conductivities $\sigma^{||,\perp,\times}_{CM}$ and the merging of $\sigma^{||,\perp,\times}_{CM}(\rho_{B},M_{q})$ with $\sigma^{||,\perp,\times}_{CM}(\rho_{B},m_{q})$ at very high baryon densities can be understood by combining Eq.~\eqref{mu_Q} with Eqs.~\eqref{as5}-\eqref{as7},
\begin{eqnarray}
&&\sigma^{||,\perp,\times}_{CM}(\rho_{B},m_{q},B)=\frac{Q^{2}}{3}\tau_{c}^{||,\perp,\times} \frac{\rho_{B}}{\sqrt{(3\pi^{2}\rho_{B})^{2/3}+m^{2}_{q}}},\label{R1}\\
&&\sigma^{||,\perp,\times}_{CM}(\rho_{B},M_{q},B)=\frac{Q^{2}}{3}\tau_{c}^{||,\perp,\times} \frac{\rho_{B}}{\sqrt{(3\pi^{2}\rho_{B})^{2/3}+M^{2}_{q}}}\label{R2}.
\end{eqnarray} 
Since $M_{q}> m_{q}$ and the difference between them decreases as $\rho_{B}$ increases, we get from Eq.~\eqref{R1}-\eqref{R2} the pattern which is observed in the left panel of Fig.~\eqref{sigma_rho} for the classical components of conductivity. The quantized parallel conductivity $\sigma^{||}_{QM}(\rho_{B},m_{q})$ (red solid line) increases with $\rho_{B}$ and aligns with the $\sigma^{||}_{CM}(\rho_{B},m_{q})$ (red dashed line) at high baryon density $\rho_{B}\sim 4\rho_{0}$. Similarly does the components $\sigma^{||}_{CM}(\rho_{B},M_{q})$ (blue dashed line) and $\sigma^{||}_{QM}(\rho_{B},M_{q})$ (blue solid line) computed considering the constituent quark mass $M_q$. The oscillating pattern in quantized conductivities $\sigma_{QM}^{||,\perp,\times}$ is an outcome of the Landau quantization with kinks corresponding to the addition of each new Landau level. This has been previously discussed in similar contexts in Refs. \cite{Satapathy:2021cjp,Satapathy:2021wex,Dey:2021fbo,Bandyopadhyay:2023lvk,Satapathy:2022xdw}.

On the other hand, the resistivity components $\rho^{||,\times,\perp}(\rho_{B},M_{q} \text{ or }m_{q}, 5 m_{\pi}^{2})$ are presented in the right panel of Fig.~(\ref{sigma_rho}) showing trends consistent with Eqs.~\eqref{as18}-\eqref{as22}. The parallel and perpendicular components of the classical resistivity $\rho^{||,\perp}_{CM}(\rho_{B},m_{q})$ (red dashed line) and $\rho^{||,\perp}_{CM}(\rho_{B},M_{q})$ (blue dashed line) decrease with increasing $\rho_{B}$, and the components with the constituent quark mass $M_{q}$ lies above the ones with current quark mass $m_{q}$ since $M_{q}>m_{q}$. These behaviours can be easily verified by resorting to Eqs.~\eqref{as18} and \eqref{mu_Q}, which give,
\begin{eqnarray}
&&\rho^{||,\perp}_{CM}(\rho_{B},m_{q},B)=\frac{1}{3Q^{2}\tau_{c}}\frac{\sqrt{(3\pi^{2}\rho_{B})^{2/3}+m^{2}_{q}}}{\rho_{B}},\label{R3}\\
&&\rho^{||,\perp}_{CM}(\rho_{B},M_{q},B)=\frac{1}{3Q^{2}\tau_{c}}\frac{\sqrt{(3\pi^{2}\rho_{B})^{2/3}+M^{2}_{q}}}{\rho_{B}}\label{R4}.
\end{eqnarray}
The classical Hall resistivity $\rho^{\times}_{CM}$ also decreases with $\rho_{B}$, consistent with Eq.~\eqref{as19}. Interestingly, the mass modification of the quarks at finite baryon density (constituent quark mass) has no effect on $\rho^{\times}_{CM}$ as the expression \eqref{as19} can be written in terms of baryon density as $\rho^{\times}_{CM}=B/3Q\rho_{B}$, independent of the quark mass. This results in the blue and red dashed curves coinciding in the bottom right panel. 
The curves of $\rho^{\times}_{QM}$ with the current quark mass and constituent quark mass also coincide in agreement with expression ~\eqref{as22} rewritten in terms of baryon density $\rho_B$ as,
\begin{eqnarray}\label{eq_hall}
&& \rho_{QM}^{\times}(\rho_{B},B)=\frac{2\pi^{2}}{3 |Q|^{3}B}\left(\sum_{l=0}^{l_{max}}l\alpha_{l}\frac{1}{{\sqrt{(3\pi^{2}\rho_{B})^{2/3}-2l|Q|B}}}\right)^{-1},\label{R5}
\end{eqnarray}
which again is independent of the quark mass. 

\begin{figure*}
	\centering
	\includegraphics[width=0.85\textwidth]{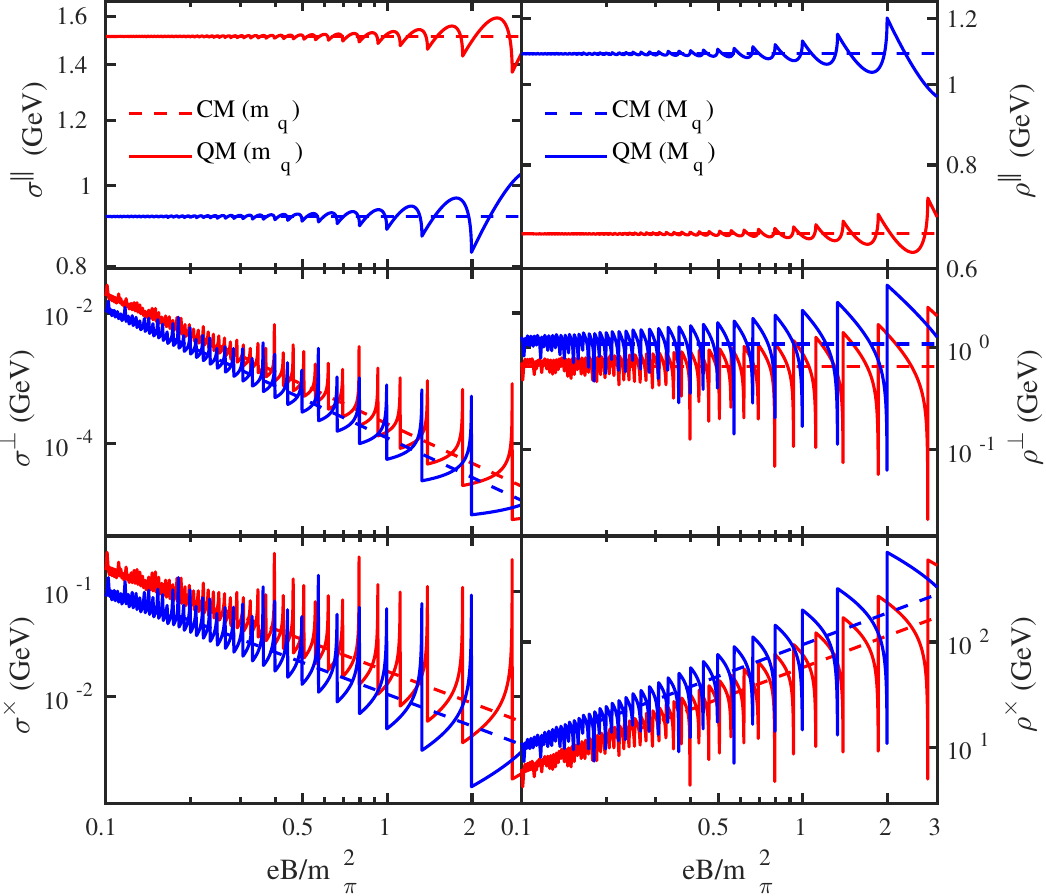}
	\caption{The variation of different components of electrical conductivity (left panel) and resistivity (right panel) as a function of magnetic field $B$ at a constant quark chemical potential $\mu_{q}=0.38$ GeV.}\label{sigma_B}
\end{figure*}
\begin{figure*}
	\centering	\includegraphics[width=0.85\textwidth]{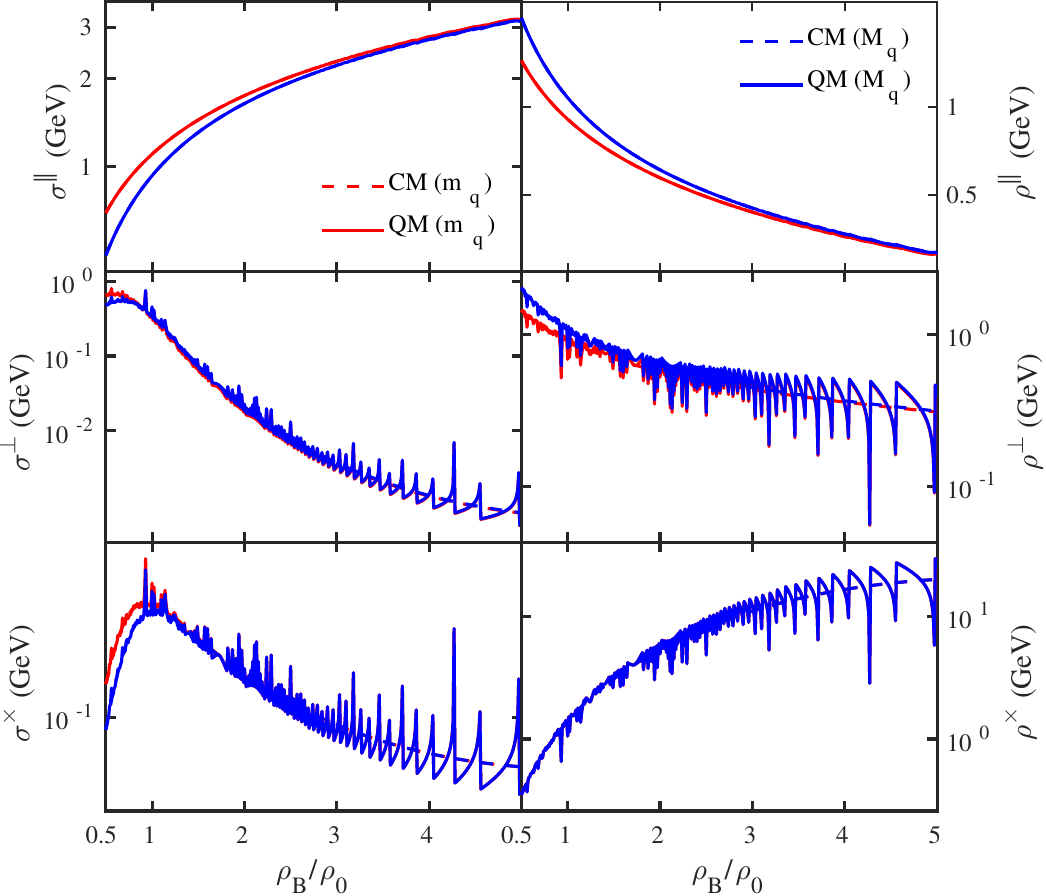}
	\caption{The variation of different components of electrical conductivity (left panel) and resistivity (right panel) as a function of scaled baryon density $\rho_{B}/\rho_{0}$ expected in the interior of a neutron star where the magnetic field varies as a function of baryon density as given in Eq.~\eqref{B}.}\label{sigma_B_rho}
\end{figure*}

In Fig.~(\ref{sigma_B}) we have shown the variation of both classical and quantized conductivities (left panel) and resistivities (right panel) as a function of $B$-field at a constant quark chemical potential $\mu_{q}=0.38$ GeV. This corresponds to a baryon density of $1.6\rho_0$ for the current quark mass $m_q$, and a baryon density of $0.75\rho_0$ for a constituent quark mass of $M_q = 0.239$ GeV. The classical parallel components of the conductivity $\sigma^{||}_{CM}(m_{q})$ (red dashed line) and $\sigma^{||}_{CM}(M_{q})$ (blue dashed line) 
remain constant as they are independent of the magnetic field. On the other hand, the classical perpendicular conductivities $\sigma^{\perp}_{CM}(m_{q},B)$ and $\sigma^{\perp}_{CM}(M_{q},B)$ decrease with increasing magnetic field. This can be understood from Eq.~\eqref{as6}, where one observes that $\sigma^{\perp}_{CM}$ is dependent on $B$-field only through the effective relaxation time  $\tau^{\perp}_{c}=\tau_{c}/(1+(\tau_{c}/\tau_{B})^{2})=\tau_{c}/(1+(QB\tau_{c}/\mu_{q})^{2})$, which decreases with increasing magnetic fields. In addition, we observe that the classical Hall conductivities $\sigma^{\times}_{CM}(m_{q},B)$ and  $\sigma^{\times}_{CM}(M_{q},B)$ first increase to a maximum and then decrease.
This is evident in Eq.~\eqref{as7}, where the expression of $\sigma_{CM}^{\times}$ is dependent on $B$ via the effective relaxation time $\tau^{\times}_{c}=\tau_{c} (\tau_{c}/\tau_{B})/(1+(\tau_{c}/\tau_{B})^{2})$. With $y=\tau_{c}/\tau_{B}=QB\tau_{c}/\mu_{q}$ we have, $\tau^{\times}_{c}=\tau_{c}~y/(1+y^{2})$.
The factor $y/(1+y^{2})$ increases with $y$ and has a maximum value of $1/2$ at  $y=1$ (or when $\tau_{c}=\tau_{B}$) and then decreases to zero as $y\rightarrow \infty$. Once we consider the constituent quark mass, the conductivity components (blue dashed curves) are significantly reduced in every direction compared to the current quark mass values (red dashed curves). 

The variation of quantized conductivities with the magnetic field show oscillations with kinks similar  to those observed for $QM$ curves when plotted against $\rho_B$. Virtually, the oscillation interval decreases as one moves from high to low magnetic fields and low to high densities, suggesting a link between distinct oscillation intervals and the quantization of conductivity.  Broadly speaking, high magnetic fields and low density regions belong to the quantum domain. At the same time, continuous phase space integrals can give reasonably accurate predictions in the classical domain of low magnetic field and high density. One can use the technique of $\zeta$-regularization, as given in \cite{Dey:2021fbo}, to eliminate the oscillating pattern and obtain a continuous envelope to the quantum curves. Merging of $CM$ and $QM$ curves in the high density and low magnetic field domains can be more explicit in such a plot.

The variation of the resistivity components $\rho^{||,\perp,\times}$ as a function of $B$ shown in the right panel of Fig.~(\ref{sigma_B}) depict almost reciprocal behaviours to that of conductivity with few changes as expected from Eqs.~\eqref{as18}-\eqref{as22}. The resistivity components $\rho^{||,\perp,\times}(M_{q},B)$ are enhanced compared to $\rho^{||,\perp,\times}(m_{q},B)$. We see the components $\rho^{||,\perp}_{CM}$ remain constant as they are independent of $B$, while $\rho^{\times}_{CM}$ increases linearly with $B$, \textit{i.e.,} $\rho^{\times}_{CM} \propto B$, which is traditionally associated with the classical Hall effect. 

In the strong magnetic field limit the Hall conductivity is dominated by the contribution from the lowest Landau level and vanishes at $l=0$. This has been discussed previously in Refs.~\cite{Hattori:2016lqx,Dey:2021fbo,Bandyopadhyay:2023lvk} where it is shown using the quantum field theoretical approach that the quark current has zero component in the transverse direction in the LLL approximation. Physically, this is because Landau level states move only along one dimension. The transverse current or transverse motion starts beyond the LLL.  Our results in Eq. (32) align well with this description.

So far, we discussed the dependence of electrical conductivity and resistivity on the baryon density and magnetic field. We observe Shubnikov-de Haas (SdH) type oscillations in the conductivity and resistivity components as a function of baryon density and magnetic fields. Quantization of Hall conductivity and resistivity are also observed. 
In condensed matter physics, SdH oscillations and QHE \cite{shoenberg2009magnetic,PhysRevLett.103.157002,PhysRevLett.105.247002,kartsovnik2011fermi,PhysRevLett.108.216803} are well-studied phenomena (typically with $\mu\sim$ eV and $T\sim$ meV, \textit{i.e.,} $\mu/T \gg 1$) in the limit $T\rightarrow 0$. Similarly, in the high energy regime we expect to observe these phenomena in the $T\rightarrow 0$ limit  (with $\mu\sim$ GeV and $T\sim$ MeV, \textit{i.e.,} $\mu/T \gg 1 $) attainable in dense cold nuclear matter like the interior of a neutron star. 

As we move from the surface to the core of a neutron star, the density and the magnetic field increases gradually. Following the parametrization in \cite{Bandyopadhyay:1997kh}, we can write the variation of magnetic field $B$ from the centre to the surface of a neutron star as a function of the baryon density $\rho_B$,
\begin{equation}\label{B}
B(\rho_B/\rho_0)=B_{surf}+B_0[1-\exp(-\beta (\rho_B/\rho_0)^\gamma)],
\end{equation} 
where the parameters chosen are $\beta$=0.01 and $\gamma$=3. $B_{surf}$ is the field at the surface of the star, equal to $10^{14}$~G, and  $B_0$ is the maximum field at the centre, taken to be $5\times 10^{18}$~G suggesting the magnetic field increases from the surface to the core of the neutron star. 
We use this relation between baryon density and magnetic field to obtain a more realistic plot of how electrical conductivity and resistivity changes as one moves from the surface to the core of a neutron star. The seemingly opposite effects of baryon density and magnetic field with regard to the quantization of electrical conductivity and resistivity can give rise to a rather unique conductivity/resistivity profile in neutron stars.

\begin{table}
	\centering
	\begin{tabular}{|c|c|c|c|c|c|c|}
			\hline
			\multicolumn{2}{|c|}{\rule{0pt}{3ex}\boldmath$\sigma^{ij} = \sigma^{ij}(\rho_B/\rho_0)$\unboldmath} & \multicolumn{3}{|c|}{\rule{0pt}{3ex}\boldmath$\sigma^{ij} = \sigma^{ij}(B) $\unboldmath } & \multicolumn{2}{|c|}{\rule{0pt}{3ex}\boldmath$\sigma^{ij} = \sigma^{ij}(\rho_B/\rho_0, B)$\unboldmath} \\[0.5em]
			\hline
			\multicolumn{2}{|c|}{\rule{0pt}{3ex}$B = 5~m_\pi^2$} & \multicolumn{3}{|c|}{\rule{0pt}{3ex}$\mu = 0.38 $ GeV} & \multicolumn{2}{|c|}{\rule{0pt}{3ex}$B = B (\rho_B/\rho_0)$} \\[0.5em]
			\hline
			\rule{0pt}{3ex}$\boldsymbol{\rho_B/\rho_0}$ & $\boldsymbol{l_{max}}$ & $\boldsymbol{eB/m_\pi^2}$ & $\boldsymbol{l_{max}(m_q)}$ &$\boldsymbol{l_{max}(M_q)}$ & $\boldsymbol{\rho_B/\rho_0}$ & $\boldsymbol{l_{max}}$  \\ [0.5em]
			\hline
			0.1 & 0 & 0.1 & 183 & 110 & 0.1 & 19126\\ 
			\hline
			0.5 & 1 & 0.5 & 36 & 22 & 0.5 & 1321\\ 
			\hline
			1 & 2 & 1 & 18 & 11 & 1 & 267\\ 
			\hline
			2 & 4 & 2 & 9 & 5 & 2 & 54\\ 
			\hline
			3 & 5 & 3 & 6 & 3 & 3 & 23\\ 
			\hline
			4 & 6 & 4 & 4 & 2 & 4 & 14 \\ 
			\hline
			5 & 7 & 5 & 3 & 2  & 5 & 10\\ 
			\hline
		\end{tabular}
		\caption{The maximum Landau level $l_{max}$ involved in the summation of the quantized components of electrical conductivity and resistivity for the results shown in Figs.~(\ref{sigma_rho}),(\ref{sigma_B}) and (\ref{sigma_B_rho}).}
		\label{table}
	\end{table}

In Fig.~(\ref{sigma_B_rho}), we sketch the components of electrical conductivity $\sigma^{||,\perp,\times}(\rho_B,B(\rho_B))$ (left panel) and resistivity $\rho^{||,\perp,\times}(\rho_B,B(\rho_B))$ (right panel) as a function of baryon density. Here, the magnetic field varies with the baryon density following the relation in Eq.~\eqref{B}. The parallel conductivity $\sigma^{||}_{CM}$ increases with the baryon density while the perpendicular $\sigma^{\perp}_{CM}$ and Hall $\sigma^\times_{CM}$  components rise to a maximum and then decreases with increasing baryon density. In Table (\ref{table}), we have tabulated the  maximum Landau level $l_{max}$ involved in the summation in Eqs.~\eqref{as10}-\eqref{as12} used to generate results shown in Figs.~\eqref{sigma_rho},\eqref{sigma_B} and \eqref{sigma_B_rho}. The quantized parallel conductivities $\sigma^{||}_{QM}(m_{q})$ and $\sigma^{||}_{QM}(M_{q})$ merges with the corresponding classical conductivities $\sigma^{||}_{CM}(m_{q})$ and $\sigma^{||}_{CM}(M_{q})$ near the surface of the star where the density  $\rho_{B}$ is less than the nuclear saturation density $\rho_{0}$. As we move towards the core (higher densities), interval of quantization increases, suggesting quantum effects like SdH and QHE will be visible. As for resistivity, both the classical components $\rho^{||,\perp,\times}_{CM}$, and the quantized components $\rho^{||,\perp,\times}_{QM}$, exhibit an approximately inverse relationship with their corresponding conductivities. Every component of quantized resistivity $\rho^{||,\perp,\times}_{QM}$ is oscillatory at higher densities and magnetic fields expected at the core of the neutron star.  Again, a quantitative estimate of the deviation in $CM$ and $QM$ curves and their merging can be obtained using $\zeta$-regularization calculations as in \cite{Dey:2021fbo}. However, in the present work it is evident from the curves in Fig.~(\ref*{sigma_B_rho}) that we can expect SdH and QHE phenomena in the neutron star environment at densities higher than, say, $2\rho_0$. Our work also suggests that a quantum treatment is essential while considering exotic scenarios like the possibility of a quark core in the interior of massive neutron stars.

\begin{figure*}
	\centering	\includegraphics[width=0.5\textwidth]{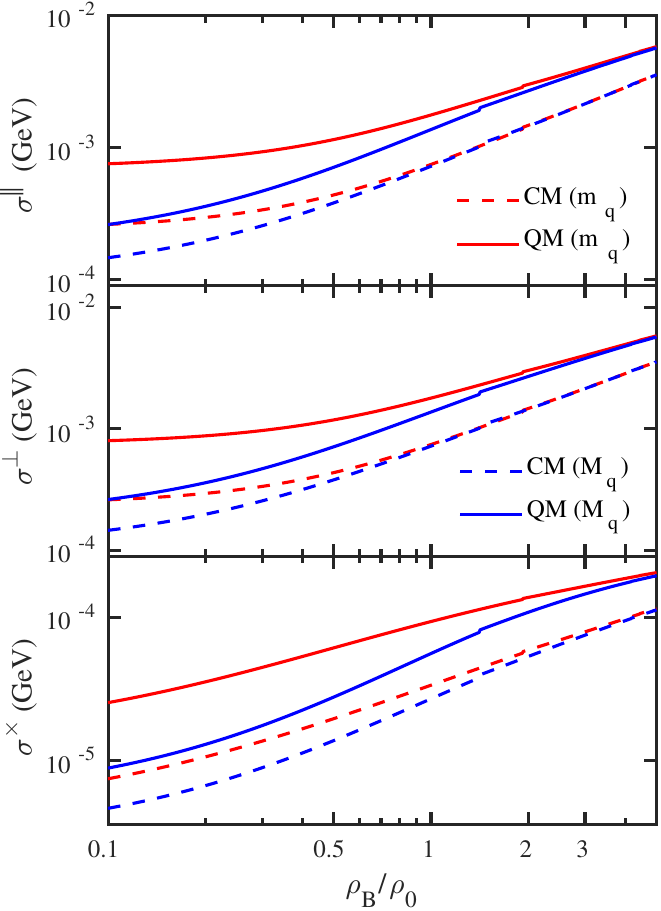}
	\caption{The variation of different components of electrical conductivity as a function of scaled baryon density $\rho_{B}/\rho_{0}$ for a constant magnetic field $eB=1~ m_{\pi}^{2}$ and temperature $T=100$ MeV.}\label{sigma_rho_T100}
\end{figure*}

Fig.~\ref{sigma_rho_T100} shows the variation of electrical conductivity components as a function of baryon density at a constant magnetic field of $eB=1~ m_{\pi}^{2}$ and temperature $T=100$ MeV as expected in the reaction zone of upcoming CBM/NICA experiments. Parallel, perpendicular and Hall components of conductivity increases with increasing baryon density. The current quark mass calculations overestimates the conductivities, particularly at low baryon densities. Also, the quantized estimates differ significantly from the classical (unquantized) estimates over the entire range of densities.

The present formalism has potential applications in heavy-ion collision experiments and neutron star observations. Electrical conductivity, defined as the electromagnetic current correlator in the zero-momentum, low-energy limit, is directly linked to the dilepton spectrum. Consequently, the anisotropic and quantized conductivity tensor can influence the angular distribution of dilepton and photon spectra in heavy-ion collisions \cite{Buividovich:2010tn,Geurts:2022xmk}. Anisotropies in electrical conductivity may enhance or suppress dilepton emission perpendicular to the reaction plane, affecting the corresponding elliptic flow. Experimental evidence for such anisotropic electrical conductivity is expected from future CBM/NICA experiments through the study of the dilepton spectrum at finite density. Modeling this bulk evolution requires magnetohydrodynamics (MHD) simulations incorporating the anisotropic and quantized conductivity tensor \cite{Benoit:2025amn,Nakamura:2022ssn,Mayer:2024dze,Mayer:2024kkv}. Similar MHD simulations are also essential for neutron star merger studies \cite{Ciolfi:2020cpf,Kiuchi:2022ubj}, where this conductivity tensor may alter simulation outcomes and indirectly impact observable astrophysical signatures.

\section{Summary}\label{IV}

Motivated by the nearly degenerate matter inside neutron stars and the reaction zone of upcoming CBM/NICA experiments, we have calculated the electrical conductivity and resistivity of matter present in such many body systems. Using the standard kinetic theory framework in relaxation time approximation, we have computed the conductivity and resistivity components along the parallel, perpendicular, and Hall directions relative to the external magnetic field.  In the present work, we obtain the constituent quark mass from the chiral effective model, which is considered a non-perturbative QCD calculation along the density axis. This external magnetic field and high density picture can be expected in the peripheral heavy-ion collisions of future facilities like CBM and NICA. It is also expected in the neutron star environment. 

The present work has systematically generated the results of the classical and quantum expressions of conductivity and resistivity components along density and magnetic field axes using straightforward current quark mass and constituent quark mass obtained from the chiral effective model. Due to the Landau quantization in the quantum expression of conductivity and resistivity tensors, they show an oscillatory or quantized behaviour along the density and magnetic field axes, often linked to SdH oscillations and QHE phenomena. They are more prominent in the low density and high magnetic field domains, which may be considered the quantum domain. We also present a more realistic calculation of conductivities and resistivities in the interior of a neutron star where the magnetic field varies with the baryon density. Using a density-dependent magnetic field profile in neutron stars, we have found a possibility of observing SdH-type oscillations in conductivity and phenomena similar to the quantum Hall effect as we move towards the core of neutron stars. This quantized pattern of the conductivity or resistivity tensor is applicable only within a limited domain of the neutron star environment, where gapless quarks in the normal (non-superconducting) phase are expected. Interestingly, as we move from the low-temperature neutron star regime to the higher-temperature conditions relevant to CBM or NICA matter, this quantization pattern disappears --- although signatures of Landau quantization and non-perturbative effects persist.

\section{Acknowledgement}
AD and DRM gratefully acknowledge the Ministry of Education (MoE), Government of India, for the financial support.  SG acknowledges that this work is partially supported by the Board of Research in Nuclear Sciences (BRNS) and the Department of Atomic Energy (DAE), Government of India, under Grant No. 57/14/01/2024-BRNS/313. The authors acknowledge the inputs and suggestions of Prof. Hiranmaya Mishra. AD, SD and DRM thank Jayanta Dey and Aritra Bandyopadhyay for valuable discussions. SG acknowledges the insightful discussion with Naosad Alam.

\bibliography{ref}

\end{document}